\documentclass[aps,showpacs,showkeys]{revtex4}

\usepackage{graphicx}
\usepackage{amsmath}
\usepackage{amssymb}

\begin{document}


\title{Effective action approach to the Leggett's mode in
two-band superconductors}

\author{S.G.~Sharapov$^{1}$}
\thanks{On leave of absence from Bogolyubov
         Institute for Theoretical Physics, Kiev, Ukraine}%
\email{Sergei.Sharapov@unine.ch}
\author{V.P.~Gusynin$^2$}
\email{vgusynin@bitp.kiev.ua}
\author{H.~Beck$^1$}%
\email{Hans.Beck@unine.ch}
\homepage{http://www.unine.ch/phys/theocond/}

\affiliation{
        $^1$Institut de Physique,
        Universit\'e de Neuch\^atel, 2000 Neuch\^atel, Switzerland\\
        $^2$Bogolyubov Institute for Theoretical Physics,
        Metrologicheskaya Str. 14-b, Kiev, 03143, Ukraine}

\date{\today}

\begin{abstract}
We investigate a collective excitation (Leggett's mode)
corresponding to small fluctuations of the relative phase of two
condensates in two-band superconductor using the effective ``phase
only'' action. We consider the possibility of observing Leggett's
mode in MgB$_2$ superconductor and conclude that for the known at
present values of the two-band model parameters for MgB$_2$
Leggett's mode arises above the two-particle threshold.

\end{abstract}




\maketitle

\section{Introduction}

The problem of collective modes in superconductors is almost as
old as the microscopic theory of superconductivity. Bogolyubov
\cite{Bogolyubov:1959:book} and Anderson \cite{Anderson:1958:PR}
discovered that
density oscillations can couple to oscillations of the phase of
the superconducting order parameter via the pairing interaction.
In a neutral system these collective soundlike oscillations are
called {\it Bogolyubov-Anderson-Goldstone\/} (BAG) mode. In a
charged system the frequency of this mode is pushed up to the
plasma frequency due to the long-range Coulomb interaction
\cite{Martin:1969:book}.

Despite the fact that the basic physics of the BAG and plasma modes in
superconductors is well understood many years ago, it appears that
a more modern approach provides even better insight to the origin
and universal character of these modes allowing also to tackle less
settled questions. A main idea beyond this approach is rather simple:
since the collective modes present usually low energy degrees
of freedom, to study them it is sufficient to have only an
effective Lagrangian (or action) that describes low-frequency,
low-wavelength dynamics of the phase $\theta$ of the pairing field
instead of working with the original Hamiltonian of the system.
(Note that although plasmons are in general
high-energy excitations, they still can be treated using an effective
theory \cite{Nagaosa:1999:book}.)

The most convenient way of deriving such a Lagrangian is to change the variables
(see, for example, \cite{Nagaosa:1999:book}) for the complex pairing field
$\Phi \to \Delta \exp(i \theta)$ with real modulus $\Delta$ and phase $\theta$.
The simultaneous transformation for the fermi field
$\psi_{\sigma} \to \chi_{\sigma} \exp(i \theta/2)$
allows from the beginning to separate in the Hamiltonian
the only phase degree of freedom relevant for the effective
theory.
In the theory of superconductivity
this replacement of the variables that resembles a gauge transformation
has, in fact, a long history and it was probably
firstly used in \cite{Ambegaokar:1984:book} and then in
\cite{Ramakrishnan:1989:PSc}
(see also Refs. in the review \cite{Loktev:2001:PRep}). More recently
this transformation was used by many authors. For example, in
\cite{Aitchison:1995:PRB} it was used to study the problem of the
Galilean invariance of the effective Lagrangian at $T=0$. The
finite temperature time-dependent effective actions for the phase
field in $s$-wave \cite{Aitchison:2000:PRB} and $d$-wave
\cite{Sharapov:2001:PRB}
{\it neutral} superconductors were derived addressing an old
problem of time-dependent generalization of the GL theory.
It is also convenient to investigate the
plasma mode within this formalism taking into account the
long-distance Coulomb repulsion between electrons
\cite{Nagaosa:1999:book,Ramakrishnan:1989:PSc} (see also more recent papers
\cite{Randeria:2000:PRB,Benfatto:2001:PRB}).

Besides these rather commonly known modes there are also other
collective modes such as {\it Carlson-Goldman\/} mode
\cite{Goldman:1975:PRL} which can appear when the Coulomb
interaction gets screened and {\it Leggett's\/} mode
\cite{Leggett:1966:PTP} which is present in two-band
superconductors. Recently  Carlson-Goldman mode has  been
studied in $d$-wave superconductors \cite{Sharapov:2002:PRB} and
colour superconductivity \cite{Gusynin:2001:PRD} using the
approach based on the modulus-phase variables, but to the best of
our knowledge Leggett's mode has not been considered yet using
this transformation. Thus it would be interesting and instructive
to apply this method to obtain Leggett's mode.

Physically Leggett's mode \cite{Leggett:1966:PTP} is a collective
excitation corresponding to small fluctuations of the relative phase
of two condensates in a clean two-band superconductor.
There are some indications that MgB$_2$ superconductor with 
$T_c = 39 \mbox{K}$ \cite{Nagamatsu:2001:Nature}
discovered about one year ago is an example of such a two-band system. 
Thus the main results of the present paper are two-fold and can be summarized 
as follows.\\
\noindent i) Leggett's mode is obtained using the modulus-phase
variables in the path integral formalism;\\
\noindent
ii) It is considered whether this mode can be observed in MgB$_2$ superconductor.

The paper is organized as follows. In Sec.~\ref{sec:model} we
introduce the two-band model and represent the partition function
of the system using the Hubbard-Stratonovich transformations for
pairing and Coulomb parts of the interactions. In
Sec.~\ref{sec:phase} we express the effective thermodynamical
potential of the system in the modulus-phase variables and obtain
the system of equations for the superconducting gaps. Then we
separate the part which describes the collective phase modes. In
Secs.~\ref{sec:neutral} and \ref{sec:charged} we derive the
dispersion law for Leggett's mode in neutral and charged systems,
respectively. Sec.~\ref{sec:MgB} we investigate the possibility of
observing Leggett's mode in MgB$_2$ superconductor. We conclude in
Sec.~\ref{sec:conclusion} with a discussion and summary of our
results.

\section{Model two-band Hamiltonian and Hubbard-Stratonovich transformations}
\label{sec:model}
Let us consider the following action (in our notations
the functional integral is expressed via $e^{S}$)
\begin{equation}
\label{S}
S = - \int_{0}^{\beta} d \tau \left[\sum_{i=1}^{2}\sum_{\sigma} \int
d^2r \psi_{i \sigma}^{\dagger}(\tau, {\bf r})
\partial_{\tau}  \psi_{i \sigma}(\tau, {\bf r}) + H(\tau) \right]\,,
\qquad \mathbf{r} = (x,y,z)\,,
\qquad \beta \equiv \frac{1}{T}\,,
\end{equation}
where the Hamiltonian $H(\tau)$ is
\begin{equation}
\label{Hamilton}
\begin{split}
H(\tau) & = \sum_{i=1}^{2}
\sum_{\sigma} \int d^2 r \psi_{i \sigma}^{\dagger}(\tau, {\bf r})
[\varepsilon_i (- i \nabla ) - \mu]
\psi_{i \sigma}(\tau, {\bf r})      \\
& - \frac{1}{2} \sum_{i,j =1}^{2}
\sum_{\sigma} \int d \mathbf{r}_1 \int d \mathbf{r}_2
\psi_{i \sigma}^{\dagger}(\tau, {\bf r}_2) \psi_{i \bar{\sigma}}^{\dagger}(\tau,
{\bf r}_1) V_{i j}({\bf r}_1; {\bf r}_2) \psi_{j \bar{\sigma}}(\tau, {\bf r}_1)
\psi_{j \sigma}(\tau, {\bf r}_2)\\
& + \frac{1}{2}  \int d \mathbf{r}_1 \int d \mathbf{r}_2
\left(\sum_{i=1, \sigma}^{2} \psi_{i \sigma}^{\dagger}(\tau, {\bf
r}_1) \psi_{i \sigma}(\tau, {\bf r}_1) -n  \right) V_{c}({\bf r}_1
- {\bf r}_2) \left(\sum_{j = 1, \sigma^{\prime}} \psi_{j
\sigma^{\prime}}^{\dagger}(\tau, {\bf r}_2) \psi_{j
\sigma^{\prime}}(\tau, {\bf r}_2) -n  \right) \,.
\end{split}
\end{equation}
Here  $\psi_{i \sigma}(\tau, {\bf r})$ is a fermion field with the spin
$\sigma= \uparrow, \downarrow$, $\bar{\sigma} \equiv - \sigma$,
$i, j =1,2$ is the band index, $\varepsilon_{i}(\mathbf{k}) = \mathbf{k}^2/2 m_i$
is the dispersion law in $i$-th band with the effective mass of the carriers $m_i$,
$\tau$ is the imaginary time and
$V_{i j}({\bf r}_1; {\bf r}_2) = V_{i j} \delta (\mathbf{r}_1 - \mathbf{r}_2)$
is an attractive short-range potential,
$ V_{c}({\bf r}_1 - {\bf r}_2)$
is the long range Coulomb interaction, $n$ is the neutralizing
background charge density. Throughout the paper we call the superconducting system
{\em neutral} if the last term of Eq.~(\ref{Hamilton}) is omitted and {\em charged}
if this term is taken into account. Even in the latter case the whole superconductor
remains, of course, neutral due to the neutralizing ionic background.
Throughout the paper $\hbar = k_B = 1$ units are chosen.

Introducing Nambu variables
\begin{equation}
\label{Nambu.variables}
\Psi_i(\tau, \mathbf{r}) = \left( \begin{array}{c}
\psi_{i \uparrow}(\tau, \mathbf{r}) \\
\psi_{i \downarrow}^{\dagger}(\tau, \mathbf{r})
\end{array} \right), \qquad
\Psi_i^{\dagger}(\tau, \mathbf{r}) = \left( \begin{array}{cc}
\psi_{i \uparrow}^{\dagger}(\tau, \mathbf{r})
\quad \psi_{i \downarrow}(\tau, \mathbf{r})
\end{array} \right)
\end{equation}
we rewrite the action as a sum
\begin{equation}\label{S.Nambu}
S = S_0 + S_{pair} + S_{C}
\end{equation}
of free
\begin{equation}
S_0 = - \int_{0}^{\beta} d \tau \int d \mathbf{r} \sum_{i=1}^{2}
\Psi_{i}^{\dagger}(x) [ \hat{I} \partial_\tau +
\tau_3 (\varepsilon_i (-i \nabla) -\mu)] \Psi_{i}(x),
\end{equation}
pairing
\begin{equation}
S_{pair} =  \int_{0}^{\beta} d \tau \int d \mathbf{r} \sum_{i,j=1}^{2}
V_{ij} \Psi_{i}^{\dagger} \tau_{+} \Psi_{i} (x)
\Psi_{j}^{\dagger} \tau_{-} \Psi_{j} (x),
\end{equation}
and Coulomb
\begin{equation}
S_{C} = - \frac{1}{2} \int_{0}^{\beta} d \tau \int d \mathbf{r}_1
\int d \mathbf{r}_2 \left(\sum_{i=1}^{2} \Psi_{i}^{\dagger} (x)
\tau_3 \Psi_{i} (x) - n \right) V_c( \mathbf{r}_1 - \mathbf{r}_2)
\left(\sum_{j=1}^{2} \Psi_{j}^{\dagger} (x) \tau_3 \Psi_{j} (x) -
n \right)
\end{equation}
parts. Here $\tau_{\pm} = (\tau_1 \pm i \tau_2)/2$,
$\tau_{\lambda}$ ($\lambda =1,2,3$) are Pauli matrices.

The easiest way to treat $S_{pair}$ is to introduce Hubbard-Stratonovich fields
$\Phi_i$ for each band, so that
\begin{equation}
\begin{split}
S_{pair} (\Phi_i, \Phi_i^{\ast}, \Psi_i, \Psi_i^{\dagger})
=  \int_{0}^{\beta} d \tau \int d \mathbf{r}
&\left[ - g_{11} |\Phi_{1}(x)|^2  - g_{22} |\Phi_{2}(x)|^2  +
g_{12} (\Phi_{1}^{\ast}(x) \Phi_{2} (x) +  \Phi_{2}^{\ast}(x) \Phi_{1}(x) )
\right. \\
& \left.
+ \sum_{i =1}^{2} g_{ii} (\Phi_{i}^{\ast}(x)
\Psi_{i}^{\dagger} \tau_{-} \Psi_{i} (x)
+  \Phi_{i}(x) \Psi_{i}^{\dagger}(x) \tau_{+} \Psi_{i} (x)) \right],
\end{split}
\end{equation}
where the coupling constants $g_{ij}$ are expressed in terms
of the original constants $V_{ij}$
\begin{equation}
g_{11} = V_{11} \left(1 - \frac{V_{12}^{2}}{V_{11} V_{22}} \right), \quad
g_{22} = V_{22} \left(1 - \frac{V_{12}^{2}}{V_{11} V_{22}} \right), \quad
g_{12} = V_{12} \left(1 - \frac{V_{12}^{2}}{V_{11} V_{22}} \right).
\end{equation}
For the repulsion part only one Hubbard-Stratonovich field
$\varphi$ is necessary:
\begin{equation}
\begin{split}
& S_{C} (\varphi,  \Psi_i, \Psi_i^{\dagger}) \\
& = \int d \tau \int d \mathbf{r}_1 \int d \mathbf{r}_2 \left[
- \frac{1}{2}e \varphi (\tau, \mathbf{r}_1) V_c^{-1}(\mathbf{r}_1
- \mathbf{r}_2) e \varphi(\tau, \mathbf{r}_2) + i e \varphi(\tau,
\mathbf{r}_1) \left(\sum_{i=1}^{2}\Psi_{i}^{\dagger}(x) \tau_{3}
\Psi_{i} (x) - n \right) \delta(\mathbf{r}_1 - \mathbf{r}_2)
\right].
\end{split}
\end{equation}
Thus the partition function of the system can be presented as
\begin{equation}
Z =  \int \mathcal{D} \Phi_{i} \mathcal{D} \Phi_{i}^{\ast}
\mathcal{D} \varphi \mathcal{D} \Psi_{i}  \mathcal{D} \Psi_{i}^{\dagger}
\exp[ S_{0} (\Psi_i, \Psi_i^{\dagger}) +
S_{pair} (\Phi_i, \Phi_i^{\ast}, \Psi_i, \Psi_i^{\dagger}) +
S_{C} (\varphi,  \Psi_i, \Psi_i^{\dagger})].
\end{equation}

\section{Effective potential in the modulus-phase variables}
\label{sec:phase}

\subsection{Modulus-phase variables}

The modulus-phase variables $\Delta_i$  and $\theta_i$ in two bands
are introduced exactly as discussed in \cite{Loktev:2001:PRep}:
\begin{equation}\label{phase}
\Phi_i (\tau, \mathbf{r}) = \Delta_i (\tau, \mathbf{r})
\exp[i \theta_i (\tau, \mathbf{r})], \qquad
\Psi_{i}(\tau, \mathbf{r}) =
\begin{pmatrix}
e^{i \theta_i (\tau, \mathbf{r})/2} & 0 \\
0 & e^{-i \theta_i (\tau, \mathbf{r})/2}
\end{pmatrix} \Upsilon_i (\tau, \mathbf{r}),
\end{equation}
making the moduli $\Delta_i$ of the pairing field real.
We note that since we restricted our analysis to $T < T_c$ in zero
external field, so that no vortices are present,
the phases $\theta_i(\tau, \mathbf{r})$ are non-singular,
so that the transformations (\ref{phase}) do not change the magnetic field
``seen'' by quasiparticles. Now absorbing $g_{11}$ and
$g_{22}$ in $\Delta_1$ and $\Delta_2$
and integrating out neutral fermions $\Upsilon_i$ we obtain
(see e.g. \cite{Gusynin:1999:JETP,Loktev:2001:PRep,Sharapov:2002:PRB})
\begin{equation}
Z = \int \Delta_{i}\mathcal{D} \Delta_{i}
\mathcal{D} \theta_{i} \mathcal{D} \varphi
\exp[- \beta \Omega (\Delta_i, \theta_i, \varphi)]
\end{equation}
with the effective thermodynamical potential
\begin{equation}
\begin{split}
\beta \Omega (\Delta_i, \theta_i, \varphi) & =
\int_{0}^{\beta} d \tau d \mathbf{r}
\left[\frac{\Delta_1^2}{g_{11}}  +  \frac{\Delta_2^2}{g_{22}}  -
\frac{2 g_{12}}{g_{11} g_{22}} \Delta_1 \Delta_2 \cos(\theta_1 - \theta_2)
\right] \\
& +  \int_{0}^{\beta} d \tau d \mathbf{r}_1  d \mathbf{r}_2
\left[ \frac{1}{2} e \varphi(\tau,\mathbf{r}_1)
V_c^{-1}(\mathbf{r}_1, \mathbf{r}_2) e \varphi(\tau,\mathbf{r}_2)
+ i e \varphi(\tau, \mathbf{r}_1) n \delta(\mathbf{r}_1 - \mathbf{r}_2)
\right]  - \mbox{Tr Ln} G_1^{-1} - \mbox{Tr Ln} G_2^{-1},
\end{split}
\end{equation}
where the Green's function
\begin{equation}
G_i^{-1}  \equiv \mathcal{G}_i^{-1} - \Sigma_i (\partial \theta_i)
=  -\hat{I} \partial_{\tau} +
\tau_{3} \left(\frac{\nabla^{2}}{2 m_i} + \mu \right) +
\tau_{1} \Delta_i (\tau,  {\bf r}) -  \Sigma_i (\partial \theta_i)
                          \label{Green.fermion.phase}
\end{equation}
with
\begin{equation}
\Sigma_i(\partial \theta_i) \equiv \tau_{3} \left[\frac{i
\partial_{\tau} \theta_i}{2} - i e \varphi(\tau, \mathbf{r}) +
\frac{(\nabla \theta_i)^{2}}{8 m_i} \right] - \hat{I}
\left[\frac{i \nabla^{2} \theta_i}{4 m_i} + \frac{i \nabla
\theta_i(\tau,  {\bf r}) \nabla}{2 m_i} \right]
                          \label{Sigma}
\end{equation}
that depends only on the time and space derivatives of $\theta_i$, but not on
the phase $\theta_i$ itself.
Then we can represent $\Omega$ as the sum
\begin{equation}
\Omega(\Delta_i, \theta_i, \varphi)  \simeq
\Omega _{\rm kin} (\Delta_i, \partial \theta_i, \varphi) +
\Omega _{\rm pot} (\Delta_i, \theta_i, \varphi),
\label{kinetic.phase+potential}
\end{equation}
where
\begin{equation}
\Omega _{\rm kin} (\Delta_i, \partial \theta_i)
 = T\, \sum_{i=1}^{2}
\mbox{Tr} \sum_{n=1}^{\infty}
\left. \frac{1}{n} (\mathcal{G}_i \Sigma_i)^{n} \right|_{\Delta_i = const}
\label{Omega.Kinetic.phase}
\end{equation}
is the sum of the energies of the phase fluctuations in each band
and
\begin{equation}
\begin{split}
\Omega _{\rm pot} (\Delta_i, \theta_i, \varphi)  =
& \left. \left(
\frac{\Delta_1^2}{g_{11}}  +  \frac{\Delta_2^2}{g_{22}}  -
\frac{2 g_{12}}{g_{11} g_{22}} \Delta_1 \Delta_2 \cos(\theta_1 - \theta_2) ]
- \mbox{Tr Ln} \mathcal{G}_1^{-1} -
\mbox{Tr Ln} \mathcal{G}_2^{-1} \right) \right|_{\Delta_i = const} \\
& +  \int_{0}^{\beta} d \tau d \mathbf{r}_1  d \mathbf{r}_2
\left[ \frac{1}{2} e \varphi(\tau,\mathbf{r}_1)
V_c^{-1}(\mathbf{r}_1, \mathbf{r}_2)
e \varphi(\tau,\mathbf{r}_2)
+ i e \varphi(\tau, \mathbf{r}_1) n \delta(\mathbf{r}_1 - \mathbf{r}_2)
\right].
\end{split}
\label{Omega.Potential.modulus}
\end{equation}
In $\Omega_{\rm pot}$ the most important for the appearance
of Leggett's mode term is the Josephson coupling energy of the condensates
in two bands. This term explicitly depends on the relative $\theta_1 - \theta_2$
phase of two condensates.

\subsection{The system of gap equations}

As pointed out in \cite{Leggett:1966:PTP} while $V_{11}$ and $V_{22}$
are completely fixed by the physics of the problem, the phase of $V_{12}$
is a matter of convention (corresponding to the choice of the relative
phases of the Bloch waves in the two bands). As in \cite{Leggett:1966:PTP}
we choose $V_{12}$ to be real and positive. In this case
the condition of minima of $\Omega_{pot}$ with respect to $\theta_1 - \theta_2$
gives $\theta_1 = \theta_2$, so that the system of the gap
equations $\partial \Omega_{pot}/ \partial \Delta_i =0$ has the form
\begin{equation}
\begin{split}
\begin{cases}
\Delta_1 - \frac{g_{12}}{g_{22}} \Delta_2 -
\Delta_1 g_{11} N_1 F(\Delta_1) =0, \\
\Delta_2 - \frac{g_{12}}{g_{11}} \Delta_1 -
\Delta_2 g_{22} N_2 F(\Delta_2) =0,
\end{cases}
\end{split}
\end{equation}
where $N_i = m_i p_{Fi}/(2 \pi^2)$ is the density of states (per spin) in
$i$-th band ($p_{Fi}$ is the Fermi momentum) and
\begin{equation}
F(\Delta_i) = \int_{0}^{\omega_D}
\frac{d \xi}{\sqrt{\xi^2 + \Delta_i^2}}
\tanh  \frac{\sqrt{\xi^2 + \Delta_i^{2}}}{2T}
\end{equation}
with the Debye frequency, $\omega_D$ which is for simplicity
assumed to be the same in each band.

This system of the equation can be transformed to the standard form
derived by Moskalenko \cite{Moskalenko:1959:FMM} and Suhl {\em et al.}
\cite{Suhl:1959:PRL}
\begin{equation}
\label{Suhl:eq}
\begin{split}
\begin{cases}
\Delta_1 [1 - V_{11} N_1 F(\Delta_1)] = \Delta_2 V_{12} N_2 F(\Delta_2), \\
\Delta_2 [1 - V_{22} N_2 F(\Delta_2)] = \Delta_1 V_{12} N_1 F(\Delta_1).
\end{cases}
\end{split}
\end{equation}
It appears, however, that in contrast to  Leggett's approach
\cite{Leggett:1966:PTP} there is no need of explicit use of the
system (\ref{Suhl:eq}) to obtain the spectrum of collective
excitations. In what follows we assume that the values of the
superconducting gaps $\Delta_1(T)$ and $\Delta_2 (T)$ can, in
principle, be determined from the system (\ref{Suhl:eq}) or taken
directly from the experiment.

\subsection{Effective potential for the collective modes}

Since the values of the gaps are fixed, i.e. there are no amplitude
fluctuations that is reasonable for $T \ll T_c$ it is sufficient
to consider only the part of effective potential
depending on $\theta_i$ and $\varphi$.

Following the same route as described, for example, in Refs.
\cite{Aitchison:2000:PRB,Sharapov:2001:PRB,Sharapov:2002:PRB}, we
arrive at the following frequency-momentum representation
\begin{equation}\label{matrix.potential}
\begin{split}
\beta \Omega\{\theta_i, \varphi \} & =
\frac{T}{8} \sum_{n = -\infty}^{\infty} \int \frac{d {\bf K}}{(2 \pi)^3}
\left( \varphi(-K) 4 e^2 V_c^{-1}(\mathbf{K}) \varphi(K)
+ \frac{8 g_{12}}{g_{11} g_{22}} \Delta_1 \Delta_2
(\theta_1(-K) - \theta_2 (-K)) (\theta_1(K) - \theta_2 (K))   \right.
\\
& \left. + \sum_{i =1}^{2}
\begin{bmatrix}
\theta_i(-K) \, \,   & e \varphi(-K)
\end{bmatrix}
\mathcal{M}_{i}^{-1}
\begin{bmatrix}
\theta_i(K) \\ e \varphi(K)
\end{bmatrix}
\right)\,,
\end{split}
\end{equation}
where the Josephson term was expanded up to quadratic term and
unimportant constant was dropped out.
The matrix $\mathcal{M}_i$ in (\ref{matrix.potential}) is
\begin{equation}\label{M}
\mathcal{M}_{i}^{-1} =
\begin{bmatrix}
-\Omega_n^2 {}^i\Pi_{33}(K) + {}^i\Lambda^{\alpha \beta}(K) K_{\alpha} K_{\beta}-
i \Omega_n K_{\alpha} {}^i\Pi_{03}^{\alpha}(K) -
i \Omega_n K_{\alpha} {}^i\Pi_{30}^{\alpha}(K) &
2 i \Omega_n {}^i\Pi_{33}(K) - 2 K_{\alpha} {}^i\Pi_{30}^{\alpha}(K) \\
-2 i \Omega_n {}^i\Pi_{33}(K) + 2 K_{\alpha}{}^i\Pi_{30}^{\alpha}(K) &
-4 {}^i\Pi_{33}(K)
\end{bmatrix}
\end{equation}
and we introduced short-hand notations $K = (i \Omega_n, {\bf K})$
with $\Omega_n = 2 \pi n T$ and
$\mathbf{K}$ being 3D vector (summation over dummy indices $\alpha, \beta =1,2,3$
is implied).
In  Eq.~(\ref{M})
${}^i\Lambda^{\alpha \beta} = {}^i\Lambda_0^{\alpha \beta} +
{}^i\Pi_{00}^{\alpha \beta}$
is the bare (unrenormalized by the phase fluctuations) superfluid stiffness,
where the current-current polarization function, ${}^i\Pi_{00}$,  is
\begin{equation}\label{Pi00}
{}^i\Pi_{00}^{\alpha \beta}(i\Omega_{n}, {\bf K}) \equiv T \sum_{l = -
\infty}^{\infty} \int \frac{d^3 k}{(2 \pi)^3}\,
{}^i\pi_{00}(i\Omega_{n},{\bf K}; i \omega_{l}, {\bf k})
v_{Fi\alpha}({\bf k}) v_{F i\beta}({\bf k})\,
\end{equation}
with the Fermi velocity
$v_{Fi \alpha}(\mathbf{k})  =
\partial \xi_i(\mathbf{k})/\partial k_{\alpha}|_{k = k_{Fi}}$
(here $\xi_i(\mathbf{k}) = \varepsilon_i (\mathbf{k}) - \mu$);
the density-density  polarization function, ${}^i\Pi_{33}$, is
\begin{equation}\label{Pi33}
{}^i \Pi_{33}(i\Omega_{n}, {\bf K}) \equiv T \sum_{l = -
\infty}^{\infty} \int \frac{d^3 k}{(2 \pi)^3}\,
{}^i\pi_{33}(i\Omega_{n},{\bf K}; i \omega_{l}, {\bf k})\,
\end{equation}
and the density-current polarization function, ${}^i\Pi_{03}^{\alpha}$ is
\begin{equation}
\label{Pi03}
{}^i\Pi_{03}^{\alpha}(i\Omega_{n}, {\bf K}) \equiv T
\sum_{l = - \infty}^{\infty} \int \frac{d^3 k}{(2 \pi)^3}\,
{}^i\pi_{03}(i\Omega_{n},{\bf K}; i \omega_{l}, {\bf k})
v_{Fi\alpha}({\bf k}) \,.
\end{equation}
${}^i\pi_{\lambda \kappa}$ in Eqs.~(\ref{Pi00}) - (\ref{Pi03}) is given by
\begin{equation}\label{pi}
{}^i \pi_{\lambda \kappa}(i\Omega_{n},{\bf K}; i \omega_{l}, {\bf k}) \equiv
\mbox{tr} [{}^i \mathcal{G} (i \omega_{l} + i\Omega_{n}, {\bf k} + {\bf K}/2)
\tau_{\lambda}{}^i\mathcal{G} (i \omega_{l}, {\bf k} - {\bf K}/2) \tau_{\kappa}]\,,
\qquad (\tau_{0} \equiv \hat{I})\,,
\end{equation}
where the neutral fermion Green's function coincides with
the usual Green's function of the BCS theory
\begin{equation}\label{Green.neutral.momentum}
{}^i\mathcal{G}(i \omega_{n},  {\bf k}) = - \frac{ i \omega_{n} \hat{I} +
\tau_{3} \xi_i( {\bf k}) - \tau_{1} \Delta_i} {\omega_{n}^{2} + \xi_i^{2}(
{\bf k}) + \Delta_i^{2}}\,, \qquad \omega_n = \pi (2n+1) T.
\end{equation}
${}^i\Lambda_{0}^{\alpha \beta}$ above
is the first order contribution to the superfluid stiffness:
\begin{equation}\label{Lambda0}
{}^i\Lambda_{0}^{\alpha \beta} =
\int \frac{d^3 k}{(2 \pi)^2} n_i(\mathbf{k})  m_{i}^{-1} \delta_{\alpha \beta}
= \frac{n_i}{m_i} \delta_{\alpha \beta},
\end{equation}
with
\begin{equation}\label{n(k)}
n_i(\mathbf{k}) = 1 - \frac{\xi_i(\mathbf{k})}{E_i(\mathbf{k})} \tanh
\frac{E_i(\mathbf{k})}{2T}\,, \qquad E_i(\mathbf{k}) =
\sqrt{\xi_i^2(\mathbf{k})+\Delta_i^2}\,.
\end{equation}
Writing Eq.~(\ref{matrix.potential}) we omitted the time derivative 
term linear in the phase 
(see e.g. \cite{Randeria:2000:PRB,Sharapov:2001:PRB}) which is
irrelevant for the present analysis.

For the purpose of completeness we give the explicit expressions
for the polarizations (\ref{Pi00}) - (\ref{Pi03}) in
Appendix~\ref{sec:A}. These expressions are important if, for
example, one is interested in the damping of the collective modes.
In what follows we consider these polarizations in the
hydrodynamic limit, $\Omega_n = 0$ and $\mathbf{K} \to 0$  at $T
=0$. In this case  ${}^i \Pi_{33} (0, \mathbf{0}) = - 2 N_i$,
${}^i \Lambda^{\alpha \beta} (0, \mathbf{0}) = {}^i
\Lambda_0^{\alpha \beta} = n_i/m_i = 2 N_i  v_{Fi}^{2}/3$ and
${}^i \Pi_{03}^{\alpha} (0, \mathbf{0}) =0$. Recall that $N_i$ is
the density of states per spin, while in \cite{Leggett:1966:PTP}
the density of states per particle was used. 

As was mentioned above, calculating the values 
${}^i \Pi_{33} (0, \mathbf{0})$ and
${}^i \Lambda^{\alpha \beta} (0, \mathbf{0}) $
there is no need to substitute the gap equations (\ref{Suhl:eq})
since the coupling constants $V_{ij}$ {\em do not enter\/}
${}^i \Pi_{33} (0, \mathbf{0})$ and
${}^i \Lambda^{\alpha \beta} (0, \mathbf{0})$, but enter
only in the Josephson coupling term. By contrast, the
original derivation of Leggett \cite{Leggett:1966:PTP} (see after Eq.~(3.8))
explicitly uses the system (\ref{Suhl:eq}). Note that in a more simple case
of one band system the Josephson term is absent and one immediately
gets the BAG mode without referring to the gap equation  
\cite{Aitchison:1995:PRB}.

We note also that
while ${}^i \Pi_{03}^{\alpha}$ is zero for the case considered in
the present paper, this term is crucial for the existence of 
Carlson-Goldman mode \cite{Goldman:1975:PRL,Sharapov:2002:PRB}.

Having the general representation for the thermodynamical potential
$\Omega\{\theta_i, \varphi \}$ we are ready to obtain the spectrum of collective
excitations. We start with a more simple case of a neutral superconductor.

\section{Collective excitations in a neutral superconductor}
\label{sec:neutral}

To consider neutral superconductor one can formally set $e = 0$
in (\ref{matrix.potential}), so that the terms with the
electric potential $\varphi$ disappear  from the equations
and we arrive at
\begin{equation}\label{neutral.potential}
\begin{split}
\beta \Omega\{\theta_i\} & =
\frac{T}{8} \sum_{n = -\infty}^{\infty} \int \frac{d {\bf K}}{(2 \pi)^3}
\left(   \frac{8 V_{12}}{V_{11} V_{22} - V_{12}^{2}} \Delta_1 \Delta_2
(\theta_1(-K) - \theta_2 (-K)) (\theta_1(K) - \theta_2 (K))   \right.
\\
& \left. + \sum_{i =1}^{2}
\theta_i(-K) M_{\theta i}^{-1} \theta_i(K)
\right)\,,
\end{split}
\end{equation}
where
\begin{equation}\label{M.theta}
M_{\theta i}^{-1} =
-\Omega_n^2 {}^i\Pi_{33}(K) + {}^i\Lambda^{\alpha \beta}(K) K_{\alpha} K_{\beta} -
i \Omega_n K_{\alpha} {}^i\Pi_{03}^{\alpha}(K) -
i \Omega_n K_{\alpha} {}^i \Pi_{30}^{\alpha}(K).
\end{equation}

As was already mentioned, we consider $M_{\theta_i}$ in  hydrodynamic
($\Omega_n = 0$, $\mathbf{K} \to 0$) limit at $T =0$:
\begin{equation}\label{M.theta.hydr}
M_{\theta_i}^{-1} = 2 N_i (\Omega_n^{2} + c_i^{2} K^2),
\end{equation}
where $c_i^{2} = v_{Fi}^{2}/3$ is the
velocity of the BAG mode in $i$-th band, so that
Eq.~(\ref{neutral.potential}) becomes
\begin{equation}\label{neutral.potential.matrix}
\beta \Omega\{\theta_i \}  =
\frac{T}{8} \sum_{n = -\infty}^{\infty} \int \frac{d {\bf K}}{(2 \pi)^3}
\begin{bmatrix}
\theta_1(-K) \, \,   &  \theta_2(-K)
\end{bmatrix}
\Theta^{-1}
\begin{bmatrix}
\theta_1 (K) \\  \theta_2(K)
\end{bmatrix}\,,
\end{equation}
with
\begin{equation}
\label{Theta}
\Theta^{-1} =
\begin{bmatrix}
M_{\theta_1}^{-1} + A & - A \\
- A & M_{\theta_2}^{-1} + A
\end{bmatrix}, \qquad
A \equiv \frac{8 V_{12} \Delta_1 \Delta_2}{V_{11} V_{22} - V_{12}^{2}}.
\end{equation}

Finally, solving the equation $\mbox{det} \Theta^{-1} =0$
for collective modes and making an analytical continuation
$i \Omega_n \to \omega + i 0$
we arrive at
\begin{equation}\label{modes.neutral}
\omega^2 = \frac{1}{2} \left[\omega_0^2 + (c_1^2 + c_2^2) K^2 \pm
\sqrt{\omega_0^4 + (c_1^2 -c_2^2)^2 K^4 -
2 \omega_0^2 \frac{N_1 - N_2}{N_1 + N_2} (c_1^2 - c_2^2) K^2} \right]
\end{equation}
with
\begin{equation}\label{omega0}
\omega_0^2 = \frac{N_1 + N_2}{2 N_1 N_2}
\frac{8 V_{12} \Delta_1 \Delta_2}{V_{11} V_{22} - V_{12}^{2}}.
\end{equation}
(For a direct comparison with \cite{Leggett:1966:PTP} recall that
the density of states used here is twice less.)
In the limit $K \to 0$ ($v_{Fi} K \ll \omega_0$) one obtains from
(\ref{modes.neutral})
\begin{equation}
\label{Leggett.neutral}
\begin{split}
\omega^2 = c^2 K^2, \quad
c^2 = \frac{N_1 c_1^2 + N_2 c_2^2}{N_1 + N_2} \quad &
\mbox{for} \quad ``-''; \\
\omega^2 = \omega_0^2 + v^2 K^2, \quad
v^2 =  \frac{N_1 c_2^2 + N_2 c_1^2}{N_1 + N_2}
\quad &
\mbox{for} \quad ``+''.
\end{split}
\end{equation}
The first solution of (\ref{Leggett.neutral}) corresponds to BAG mode, while
the second solution is Leggett's mode.
This  collective mode is only possible if
$\omega_0^2 > 0$. Since $V_{12} > 0$ this implies that  Leggett's mode
exists for $V_{11} V_{22} - V_{12}^2  >0 $.

\section{Collective excitations in a charged superconductor}
\label{sec:charged}

As was discussed in Introduction the long-distance Coulomb
interaction has a drastic influence on the BAG mode transforming it in
the plasma mode. Here we consider how Leggett's mode is affected
by the Coulomb interaction.
Let us rewrite action (\ref{matrix.potential}) in the hydrodynamic limit as one
matrix
\begin{equation}\label{matrix.potential.new}
\begin{split}
\beta \Omega\{\theta_i, \varphi \} & =
\frac{T}{8} \sum_{n = -\infty}^{\infty} \int \frac{d {\bf K}}{(2 \pi)^3}
\begin{bmatrix}
\theta_1(-K) \, \,    & \theta_2(-K) \, \,&  e \varphi(-K)
\end{bmatrix}
\mathcal{M}^{-1}
\begin{bmatrix}
\theta_1(K) \\ \theta_{2}(K) \\ e \varphi(K)
\end{bmatrix}\,,
\end{split}
\end{equation}
where
\begin{equation}\label{M.new}
\mathcal{M}^{-1} =
\begin{bmatrix}
M_{\theta_1}^{-1} + A & - A              & -4 i \Omega_n N_1 \\
-A               & M_{\theta_2}^{-1} + A & -4 i \Omega_n N_2 \\
4 i \Omega_n N_1 &  4 i \Omega_n N_2 & 4 (2N_1 + 2N_2 + V_{c}^{-1}(\mathbf{K}))\end{bmatrix}
\end{equation}
and $M_{\theta_1}^{-1}$ and $A$ are given by
Eqs.~(\ref{M.theta.hydr}) and (\ref{Theta}), respectively. As one
can readily see, setting $e=0$ in Eq.~(\ref{matrix.potential.new})
we reduce the action (\ref{matrix.potential.new}) with $3 \times
3$ matrix $\mathcal{M}^{-1}$ that describes the phase fluctuations
in the charged system to the action
(\ref{neutral.potential.matrix}) with $2\times 2$ matrix
$\Theta^{-1}$ that describes the collective excitations in a
neutral system.

Again the spectrum of collective excitations can be found solving
the equation $\mbox{det} \mathcal{M}^{-1} =0$. Neglecting the
inverse of Coulomb interaction in the limit $\mathbf{K} \to 0$
($V_{c}^{-1}(\mathbf{K}) =\mathbf{K}^2/(4\pi e^2) \to 0$), we
obtain
\begin{equation}
\label{Leggett.charged}
\Omega^2 = \omega_0^2 + v^2 K^2,
\qquad
v^2 = \frac{(N_1 + N_2) c_1^2 c_2^2}{N_1 c_1^2 + N_2 c_2^2}.
\end{equation}
Since the inverse Coulomb potential $\sim K^2$ is neglected, the
equation for the collective modes has the only solution describing
Leggett's mode. Comparing (\ref{Leggett.neutral}) and
(\ref{Leggett.charged}) one can see that in accordance with
\cite{Leggett:1966:PTP} the value of $\omega_0$ is insensitive to
the Coulomb interaction but the velocity $v$ is affected. It is
interesting that the value of $v$ itself does not depend on the
explicit form of $V_c(\mathbf{K})$.

Including the Coulomb interaction
$V_{c}(\mathbf{K})$ one can also check that
the equation $\mbox{det} \mathcal{M}^{-1} =0$
has a plasma mode as a solution. Indeed considering  the simplest 
$V_{12} = 0$ ($A = 0$) case we obtain the plasma frequency for in-phase
oscillations in both bands
\begin{equation}
\Omega_p^2 = 8 \pi e^2 (N_1 c_1^2 + N_2 c_2^2) =  
4 \pi e^2 \left( \frac{n_1}{m_1} + \frac{n_2}{m_2} \right).
\end{equation}

\section{Implications for M\lowercase{g}B$_2$}
\label{sec:MgB}

There are currently many indications that the recently discovered
MgB$_2$ superconductor \cite{Nagamatsu:2001:Nature} can be
described by the classical two-gap model
\cite{Moskalenko:1959:FMM,Suhl:1959:PRL} which convincingly fits
the specific heat \cite{Bouquet:2001:EPL} and penetration depth
measurements \cite{Kim:2002}. In
particular, in \cite{Liu:2001:PRL,Golubov:2002:JPCM,Barabash:2002} even the
values of the coupling constants  for the system of equations
(\ref{Suhl:eq}) for the two-gap and two-band model of MgB$_2$ were
given.

We note that one of the alternative explanations of the observed
anisotropy of the upper critical field uses a model with 
anisotropic $s$-wave pairing \cite{Posazhennikova:2002} which is
also rather close to the two-gap and two-band scenario.

To be experimentally observable Leggett's mode should have the
value of $\omega_0$ in (\ref{Leggett.charged}) well separated from
the two-particle threshold given by the smallest gap, $\Delta_1$.
Here we estimate the value of $\omega_0$ using recently suggested
values of the coupling
constants that enter the system of equations (\ref{Suhl:eq}).
Introducing the dimensionless coupling constants, $\lambda_{ij} =
N_{i} V_{ij}$ that are often used for the description of the
two-band model, we may rewrite Eq.~(\ref{omega0}) in the following
form
\begin{equation}
\omega_0^2 = \frac{\lambda_{12} + \lambda_{21}}
{\lambda_{11} \lambda_{22} - \lambda_{12} \lambda_{21}} 4 \Delta_1 \Delta_2 .
\end{equation}
Our estimates of $\omega_0$ are summarized in Table~\ref{tab1}.

\begin{table}[h]
\caption{The estimate of the frequency $\omega_0$. The values of
the superconducting $\Delta_1 = 1.8$ meV and $\Delta_2 = 6.8$ meV
from \cite{Choi:2001} are used.} \vspace{5mm}
\begin{tabular}{|l|c|c|c|c||c|c|}
\hline \hline Reference
& $\lambda_{11}$ & $\lambda_{22}$ & $\lambda_{12}$ & $\lambda_{21}$ &
$\omega_0$, meV & $\omega_0/2 \Delta_1$ \\
\hline Liu, {\em et al.}  \cite{Liu:2001:PRL} and Barabash \cite{Barabash:2002}
& 0.96    & 0.28 & 0.16   & 0.22  & 8.9 & 2.5 \\
\hline Golubov,  {\em et al.} \cite{Golubov:2002:JPCM} & 1.017 & 0.448 & 0.213  & 0.155 & 6.5  & 1.8 \\
\hline \hline
\end{tabular}
\label{tab1}
\end{table}

They show that for the values of the two-band model parameters 
known at present for the two-band model of MgB$_2$, Leggett's mode
arises above the two-particle threshold, $\omega_0 / 2 \Delta_1 >
1$, and unlikely to be observed. We do not exclude, however, that
Leggett's mode can be observed in MgB$_2$ if the values of the
coupling constants are revised, so that the interband coupling
constants $\lambda_{12}$ and $\lambda_{21}$ would become smaller
allowing $\omega_0/2 \Delta_1 < 1$. The observation of Leggett's
mode would provide an additional insight to the underlying physics
of this superconductor.

\section{Conclusion}
\label{sec:conclusion}

To conclude, we readdressed the collective excitations of the
relative phase of the two condensates in a clean two-band
superconductor using the effective ``phase action'' formalism.
This formalism has proved itself as a convenient and economic
method of the investigation of the collective modes in
superconductors. Our estimates of the lowest frequency of
Leggett's mode for MgB$_2$ show that it can hardly be observed in
this superconductor.

\section{Acknowledgments}

This work was supported by the research project 20-65045.01
by the SCOPES-project 7UKPJ062150.00/1 of the Swiss National
Science Foundation.

\appendix
\section{The expressions for polarizations $\Pi_{ij}$}
\label{sec:A}

The expressions for the polarization functions
$\Pi_{\lambda \kappa}$ (for simplicity we omit the band index $i$) are
(see e.g. \cite{Sharapov:2001:PRB,Aitchison:2000:PRB})
\begin{equation}
\label{Pi.general}
\begin{split}
& \begin{bmatrix}
\Pi_{00}^{\alpha \beta}(i \Omega_{n}, {\bf K}) \\
\Pi_{33}(i \Omega_{n}, {\bf K})
\end{bmatrix} = \\
\\ &  - \int \frac{d^3 k}{(2 \pi)^3}  \left\{ \frac{1}{2} \left(1 -
\frac{\xi_{-}\xi_{+} \pm \Delta^{2} }{E_{-}E_{+}} \right) \left[
\frac{1}{E_{+} + E_{-} + i \Omega_n} +
\frac{1}{E_{+} + E_{-} - i \Omega_n}\right] [1 - n_F(E_{-}) - n_F(E_{+})]
\right. \\
& \left. + \frac{1}{2} \left(1 + \frac{\xi_{-}\xi_{+} \pm \Delta^{2}
}{E_{-}E_{+}} \right)\left[ \frac{1}{E_{+} - E_{-} + i \Omega_n} +
\frac{1}{E_{+} - E_{-} - i \Omega_n}\right] [n_F(E_{-}) - n_F(E_{+})]
\right\}V_{\pm}(\mathbf{k})\,,\\
& \qquad \qquad \qquad V_{\pm}(\mathbf{k}) \equiv
\begin{bmatrix}
v_{F \alpha}(\mathbf{k}) v_{F \beta}(\mathbf{k}), & ``+''; \\
1, & ``-''.
\end{bmatrix}\,,
\end{split}
\end{equation}
and
\begin{equation}\label{Pi_{03}.general}
\begin{split}
& \Pi_{03}^{\alpha}(i \Omega_{n}, {\bf K}) =\\
& =  \int \frac{d^3 k}{(2 \pi)^3} \left\{ \left(\frac{\xi_{+}}{2E_{+}} -
\frac{\xi_{-}}{2E_{-}} \right) \left[ \frac{1}{E_{+} + E_{-} + i \Omega_n} -
\frac{1}{E_{+} + E_{-} - i \Omega_n}\right] [1 - n_F(E_{-}) - n_F(E_{+})]
\right. \\
& \left. + \left(\frac{\xi_{+}}{2E_{+}} + \frac{\xi_{-}}{2E_{-}} \right) \left[
\frac{1}{E_{+} - E_{-} + i \Omega_n} - \frac{1}{E_{+} - E_{-} - i
\Omega_n}\right] [n_F(E_{-}) - n_F(E_{+})] \right\}
v_{F \alpha}(\mathbf{k}),
\end{split}
\end{equation}
where $\xi_{\pm} \equiv \xi({\bf k} \pm {\bf K}/2)$, $E_{\pm} \equiv E({\bf k}
\pm {\bf K}/2)$.
One can also check that  $\Pi_{30}^{\alpha}(i\Omega_{n},{\bf K}) =
\Pi_{03}^{\alpha} (i\Omega_{n},{\bf K})$.
The second term in Eqs.~(\ref{Pi.general}) and (\ref{Pi_{03}.general})
gives the contribution of the thermally
excited quasiparticles (i.e. ``normal'' fluid component). This is
the term responsible for the appearance of the {\em Landau terms}
in the effective action \cite{Aitchison:2000:PRB,Sharapov:2001:PRB}.

\end{document}